\documentclass[a4paper,12pt]{article}

\usepackage[english]{babel}
\usepackage{csquotes}
\usepackage{hyperref}
\usepackage{graphicx}
\usepackage{array}
\usepackage{longtable}

\usepackage[backend=biber,style=apa]{biblatex}
\title{Pursuing transparency: how research organizations in Germany collect data on publication costs}
\date{June 15, 2026}

\author{
Dorothea Strecker\thanks{Humboldt-Universität zu Berlin, Berlin School of Library and Information Science, Dorotheenstr.~26, 10117 Berlin, Germany. Email: \href{mailto:dorothea.strecker@hu-berlin.de}{dorothea.strecker@hu-berlin.de}.\textbf{ORCID}: \href{https://orcid.org/0000-0002-9754-3807}{0000-0002-9754-3807}}\\
{\small IBI, HU Berlin}
 \and
Heinz Pampel\thanks{Humboldt-Universität zu Berlin, Berlin School of Library and Information Science, Dorotheenstr.~26, 10117 Berlin, Germany and Helmholtz Association, Helmholtz Open Science Office, Telegrafenberg, 14473 Potsdam, Germany. Email: \href{mailto:heinz.pampel@hu-berlin.de}{heinz.pampel@hu-berlin.de}. \textbf{ORCID}: \href{https://orcid.org/0000-0003-3334-2771}{0000-0003-3334-2771}}\\
{\small IBI, HU Berlin}\\
{\small OS Office, Helmholtz}
 \and
Jonas Höfting\thanks{Humboldt-Universität zu Berlin, Berlin School of Library and Information Science, Dorotheenstr.~26, 10117 Berlin, Germany. Email: \href{mailto:jonas.hoefting@hu-berlin.de}{jonas.hoefting@hu-berlin.de}. \textbf{ORCID}: \href{https://orcid.org/0009-0003-4466-1775}{0009-0003-4466-1775}}\\
{\small IBI, HU Berlin}
}

\begin{document}
\maketitle

\section*{Abstract}
This article presents the results of a survey conducted in 2024 among research performing organizations (RPOs) in Germany on how they collect data on publication costs. Of the 583 invitees, 258 (44.3\%) completed the questionnaire. This survey is the first comprehensive study on the recording of publication costs at RPOs in Germany.
The results show that the majority of surveyed RPOs recorded publication costs at least in part. However, procedures in this regard were often non-binding. Respondents' ratings of the reliability of the collection of data on publication costs varied by the source of publication funding. Eighty percent of respondents rated the contribution of collecting data on publication costs to shaping the open access transformation as ``very important'' or ``important.'' Yet, these data were used as a basis for strategic decisions in only 59\% of the surveyed RPOs. Moreover, most respondents considered the implementation of an information budget at their institutions by 2025 unlikely. 
We discuss the implications of these findings for the open access transformation.

\textbf{Keywords:} open access transformation, monitoring, publication costs, information budget, survey

\section{Introduction}
Open access policies have yielded multiple approaches to transforming academic publishing through alternatives to the traditional subscription model. However, there is growing criticism that these alternative models make it more difficult to keep track of financial flows to publishers and undermine cost transparency at research performing organizations (RPOs). Against this backdrop, the primary aim of this article is to present the methodology and results of a first-of-its-kind survey of RPOs in Germany on how they collect and use data on their publication-related expenditures. By way of contextualization, we preface this presentation with an overview of international and country-level initiatives and efforts to promote the open access transformation and the monitoring of publication output and costs.

\subsection{Transformation of Academic Publishing to Open Access}
Early steps towards removing barriers to scholarly literature began in the 1990s. Ever since three statements gained widespread attention in the early 2000s (the Budapest Open Access Initiative in 2002, the Bethesda Statement on Open Access Publishing in 2003, and the Berlin Declaration on Open Access to Knowledge in the Sciences and Humanities in 2003), one can speak of an "Open Access movement" \parencite{suber2012}. Eventually, institutions and funding bodies began integrating Open Access principles into their policies. In 2009, the European Heads of Research Councils (EUROHORCs) and the European Science Foundation (ESF) proposed the implementation of a common policy on open access to research results and declared: ``The aim is a system of scientific publications in which free access to all (published) scientific information is guaranteed. This involves a move toward Full Open Access'' \parencite[17]{eurohorcs2009}. The goal of the transition to open access was supported by the Council of the European Union in 2016. In what has been described as a ``dramatic statement'' \parencite{enserink2016}, the Council called on Member States, the Commission, and stakeholders ``to remove financial and legal barriers'' to ``a full scale transition towards open access'' \parencite[7]{council2016}. Although lacking a specific timeline, this goal is also supported by many RPOs and research funding organizations (RFOs) worldwide through the OA2020 initiative, which was launched by the Max Planck Society in 2016. The 158 signatories of the OA2020's ``Expression of Interest in the Large-Scale Implementation of Open Access to Scholarly Journals'' (as of November 2025) endorse the aim ``to transform a majority of today's scholarly journals from subscription to {[}open access{]} publishing''\parencite{oa2020}. This transformative approach is also supported by the RFOs that have endorsed Plan S \parencite{coalition_s2018} and that require their beneficiaries to publish their funded research publications open access.

Transforming the academic publishing system to open access ``means replacing the present reader-paid publication system with an author- or institution-paid one'' \parencite[17]{eurohorcs2009}. Therefore, a core element of the open access transformation is changing the business models of academic journals \parencite{suber2012}. From the authors' perspective, there are two primary business models for open access journals: One involves the payment of article processing charges (APCs) covered by institutions or funders; the other, known as diamond open access, does not. In addition to Open Access journals, there is also the option of self-archiving publications in repositories, known as the Green Open Access route \autocite{bjork_anatomy_2014}. The APC model has gained traction since the early 2000s. The journal \emph{PLOS Biology}, for example, adopted this model at the time of its launch in 2003 \parencite{wired2003}. The acquisition of BioMed Central by Springer Science+Business Media (now Springer Nature) in 2008 \parencite{cockerill2008} made it clear that APC-funded open access had established itself as a profitable business model, particularly in the science, technology, and medicine (STM) disciplines. However, a glance at the open access journal landscape as depicted by the Directory of Open Access Journals (DOAJ) shows that the APC-based model is not the dominant business model overall: As of July 2025, only 36.5\% of the journals listed in the DOAJ charged publication fees.\footnote{Directory of Open Access Journals: \url{https://doaj.org} (accessed July 13, 2025).} Yet, most open access journal articles are published in journals that charge APCs \parencite{crawford2019}, and commercial publishers - particularly in Asia, the United States, Canada, and Western Europe - rely predominantly on this model \parencite{bosman2021}.

Publishing in journals that charge APCs requires institutional support. In 2009, the initial signatories of the Compact for Open-Access Publishing Equity (COPE) committed to establishing ``durable mechanisms for underwriting reasonable publication charges'' for articles published by their faculty members in open access journals \parencite{noauthor2009}, thereby recognizing the APC-based model as equivalent to subscription-based models \parencite{shieber2009}. To streamline APC management and ensure centralized funding support for researchers - especially when RFOs do not reimburse APCs - RPOs have created dedicated open access publication funds \parencite{solomon2012,tananbaum_north_2014}. Since the conception of the idea, publication funds have become quite common in Germany: According to data from the oa.atlas, 368 (50 \%) of the 736 institutions listed have a journal publication fund.\footnote{oa.atlas, Open-Access-Network; accessed: 2026-06-25; filter: OA journals publication fund - yes.} These funds are often linked to institutional open access policies and require that certain eligibility criteria be met \parencite{pampel2017}. For example, some funds exclude publications in hybrid journals (i.e., subscription-based journals that offer authors the option of publishing open access for a fee) to avoid supporting ``double dipping'', ``where publishers collect money twice: once when subscriptions are paid by the universities and research organizations and once when authors are additionally charged open access fees'' \parencite{mittermaier2015}.

These new funding models have led to the development of new types of agreements between publishers and RPOs. Initially referred to as offsetting models, most notably implemented through agreements with Springer Nature in the Netherlands (2015), the United Kingdom (2015), Austria (2016), and Sweden (2016) \parencite{freedman2016}, they later evolved under the label ``Publish and Read'' (PAR) agreements. These agreements aim to integrate the costs of subscriptions and open access publishing across a publisher's portfolio. They are sometimes also referred to as ``transformative agreements,'' thus ascribing them a central role in facilitating the transformation of subscription-based journals to an open access business model (``journal flipping''). However, this objective has not yet been realized in practice \parencite{mittermaier2025}.

\subsection{Monitoring Publication Output}
Open access policies such as those mentioned above have led to the development of procedures and tools for monitoring open access publication output to track compliance. At an international level, the launch in 2017 of the Open Science Monitor, which was developed by RAND Europe and colleagues on behalf of the European Commission, marked a first major step in open access monitoring \parencite{rand_nodate}. The Open Science Monitor was updated and expanded in 2018 by a consortium led by the Lisbon Council, with Elsevier as a subcontractor, which triggered public and methodological criticism \parencite{french_open_science_nodate,moody2018a,moody2018b}. The OpenAIRE Monitor, an on-demand service based on the OpenAIRE Research Graph, which was launched in 2020 and expanded in 2022 to include institutional dashboards \parencite{pispiringas2022}, marked a further major step in open access monitoring. As of June 2025, further development of open access monitoring was planned within the framework of the project EOSC Future \parencite{oneill2022}. Since 2017, several European countries have launched national open access monitoring initiatives. Discussions on the need to monitor the development of open access began initially in Finland \parencite{olsbo2017}. Germany was likely the first to implement a fully operational national dashboard, the German Open Access Monitor (OAM), which was developed between 2018 and 2020 \parencite{mittermaier2021,barbers2021}. France followed with the Baromètre français de la Science Ouverte in 2019 \parencite{bracco2022}; Switzerland launched its national dashboard in 2022 \parencite{lib4ri2022}; and Austria developed national coordination structures \parencite{danowski2018}, culminating in the 2024 beta release of the Austrian Datahub. Institutional open access monitoring differs between RPOs and RFOs. RPOs such as Forschungszentrum Jülich \parencite{julich_nodate}, the University of Zurich \parencite{zurich2024}, and Freie Universität Berlin \parencite{duine2024} have implemented dashboards to visualize open access publication spending - the Open Access Barometer, the Open Access Monitor, and the Open Science Dashboard for Earth Sciences, respectively. RFOs rely on tools such as CHORUS \autocite{dylla2014, poynder2013}, the Europe PMC Funders’ Dashboard \autocite{pmc2020}, and OpenAIRE dashboards \autocite{brunschweiger2023}, or collect publication data as part of the reporting process. Identifying publications related to specific grants in bibliographic databases is difficult because of sparse funding information and other metadata quality issues \parencite{mugabushaka_funding_2022}.

\subsection{Monitoring Publication Costs}
The open access transformation has always been accompanied by discussions on the financial dimension of publishing. In 2015, \citeauthor{schimmer_disrupting_2015} published a white paper with an estimate of the total costs associated with academic publishing, arguing that a full transition to open access was financially feasible. In recent years, the ``big five'' commercial publishers (Elsevier, Sage, Springer Nature, Taylor \& Francis, and Wiley) have adopted the APC model \parencite{butler2023}. This has been accompanied by criticism, as APCs are increasing, and it is debatable whether they reflect the true costs of publishing research articles \parencite{butler2024b,kemp_cost_2024}.

Funding models for Open Access publishing mean that RPOs are now confronted not only with subscription fees but also with transformative agreements, APCs, and additional publication-related expenditures. The administration of financial flows from RPOs to publishers is also changing, as library budgets are no longer the only funding sources. A survey of RPOs in Germany conducted in 2018 found that library funds, research grants, and departmental budgets were used to cover publication costs \parencite{pampel2021}. A survey of Springer Nature authors on the sources of their APC funding revealed a similar picture \parencite{monaghan2020}. However, prices are reported inconsistently, and information on costs is rarely made public \parencite{kemp_cost_2024}. This fragmentation and obfuscation makes it increasingly difficult to obtain a comprehensive overview of all publication-related expenditures of RPOs.

Studies on the financial implications of the open access transformation at the level of RPOs have been conducted in the United Kingdom \parencite{research_consulting2014,swan2010,swan2012}, the United States \parencite{harlan2024,smith2016}, Germany \parencite{barbers2018,jahn2016,taubert2019}, Sweden \parencite{loven2019}, and Switzerland \parencite{cambridge2017}. In addition, the topic was addressed in a study covering the EU Member States, Norway, the United Kingdom, and Switzerland, which was commissioned by the European Commission in 2024 \parencite{european_commission2024}. Other studies have analyzed the costs associated with RFOs' support for open access publishing, for example, in Spain \parencite{alonso2024} and Germany \parencite{barbers2023,barbers2025}.

Establishing a structured approach to managing publication fees such as APCs is key to supporting researchers in publishing open access while ensuring efficient cost management. The Efficiency and Standards for Article Charges (ESAC) initiative is working on standardizing processes for handling APCs between RPOs and publishers \parencite{esac2017,geschuhn2017}. At the core of these efforts is the development of open access article workflows that enable the effective and efficient management of cost-related data. The challenges associated with establishing procedures for monitoring and processing open access publication costs have been the subject of several studies and practical case examples in Austria \parencite{pinhasi2018}, Germany \parencite{beckmann2022,eppelin2012,frick2017,schoen2024,sikora2015}, and Scotland \parencite{nixon2013}.

The disclosure of the costs of academic communication has been discussed as a measure to increase cost transparency for some years now \parencite{amsterdam2016,europe_science2017}. A recommendation issued by the European Commission in 2018 went beyond open access, stating that, ``to enhance market transparency and fair competition'' EU Member States should ensure that ``information is published about agreements between public institutions or groups of public institutions and publishers on the supply of scientific information'' \parencite[p. L134/14]{european_commission2018}.

Several initiatives are working to address the current lack of cost transparency. Notable examples include the OpenAPC initiative, which aggregates data on fees paid for open access journal articles by RPOs and RFOs \parencite{pieper2018}, price transparency requirements by RFOs\footnote{cOAlition S announces price transparency requirements:\url{https://www.coalition-s.org/coalition-s-announces-price-transparency-requirements/}}, as well as efforts to systematically collect and analyze information on APC pricing from the websites of major publishers \parencite{butler2023,haustein2024,kendall2024}. In addition, freedom of information requests have been used in Finland \parencite{lahti2018}, the United Kingdom \parencite{lawson2015}, New Zealand \parencite{wilson2014}, and Switzerland \parencite{gutknecht2018} to obtain information about RPOs' journal subscription expenditures. Although these initiatives use different approaches to achieving cost transparency, they all involve significant manual effort.

As described above, RPOs have been affected by increasingly fragmented financial flows since the beginning of the open access transition. If RPOs had a systematic overview of all income streams and expenditures relating to scientific information, they could more effectively evaluate the allocation of funds, compare their financial structures to those of other RPOs, and make informed decisions about future actions. In Germany, the idea of consolidating distributed data on all information-related income streams and expenditures within an RPO has been discussed for several years now under the label ``information budget'' \parencite{mittermaier2022,pampel2022,taubert2022,woitschach2025}. The term "refers to a financial management tool used to manage all income and expenditure attributable to the publishing and reception of literature at a research performing organization. As part of the organization’s budget, the information budget enables the management of all financial resources for services and products of scientific information." \autocite[7]{pampel2022} Whereas the collection budget of libraries historically focuses on subscription fees and more recently OA publication funds, the information budget includes other expenditures \cite{pampel2022}:
\begin{enumerate}
    \item Expenditures on subscriptions
    \item Expenditures on publication fees for gold OA
    \item Publication fee expenditures for hybrid options
    \item Expenditures on publication fees for closed access
    \item Expenditures for licensing of illustrations
    \item Expenses for consortial OA infrastructures
    \item Expenses for local OA infrastructures.
\end{enumerate}
So far, the concept has been discussed primarily in Germany, where it has also been addressed by a prominent research policy stakeholder: The German Council of Science and Humanities has recommended that research institutions in Germany establish such an information budget by 2025 \autocite[p. 77]{wissenschaftsrat2022}.
Since the fragmentation of publication-related funding streams is experienced by RPOs globally\autocite{kemp_cost_2024}, "information budget" could be a useful concept elsewhere.

\subsection{The Situation in Germany}
Open access is firmly established in Germany. In 2024, 69.5\% of the publication output of researchers at German RPO's was freely accessible.\footnote{Open Access Monitor, Forschungszentrum Jülich; accessed: November 13, 2025; filters: source - OpenAlex, release date - January 1, 2024, to December 12, 2024} Open access is also an explicit science policy goal, expressed, for example, in the joint guidelines on open access adopted by the Federal Government and the Länder in 2023 \parencite{bmbf2023}. As described above, in reaction to the increasing fragmentation of financial flows to publishers, some RPOs are trying to achieve a systematic overview of all income streams and expenditures relating to scientific information. To gain an insight into the status quo of these efforts in Germany, we conducted a survey among RPOs on how they collect data on publication costs, on the workflows and tools at their institutions, and the challenges they face. The survey is the first comprehensive study of the recording of publication costs at RPOs in Germany. Results of the survey have previously been published in German \parencite{strecker_erfassung_2025}. As they may also be internationally relevant, we are now presenting them to a wider audience.

\section{Method}
The survey was conducted in 2024 as part of OA Datenpraxis, a project funded by the German Research Foundation (DFG), which explores various aspects of data use in the context of the open access transformation \parencite{pampel2024b}. One goal of this project was to survey the status quo of recording publication costs at RPOs in Germany. The survey was conceptualized in cooperation with the DEAL Consortium (formerly Projekt DEAL) initiated by the Alliance of German Science Organizations.\footnote{About DEAL: \url{https://deal-konsortium.de/en/about-deal/rationale-and-objectives}} Representatives of the DEAL Consortium were involved in the development of the questionnaire and promoted the survey among institutions participating in the DEAL transformative agreements and through other channels.

\subsection{Survey Design}
The survey was guided by the question: How are data on publication costs currently recorded at research performing organizations in Germany? The survey was implemented as an online questionnaire.

\subsubsection{Compilation of the Mailing List}
As a first step, between June 20th 2024 and September 5th 2024, a list of research performing organizations in Germany was compiled based on existing directories. These were reviewed and adjusted where necessary, e.g. because of duplicates. The number of institutions in the final list by type is shown in Table \ref{table:verteiler}. Research performing organizations were defined as publicly funded organizations where research conducted by employed staff is a core activity of the organization. After compiling the list, contact information was researched. The primary selection criterion for contact persons was proximity to relevant processes, prioritizing Open Access officers, publication departments, and library directors. Individuals were preferred over departments or functional addresses. For more information on how the mailing list was created, see our data publication \autocite{strecker2025}.

\subsubsection{Questionnaire Development}
An initial draft of the questionnaire was developed after reviewing relevant literature, including workshop and practice reports and publications from related working groups \parencite{mittermaier2022,oan2024,schoen2024}. The draft included questions about conditions at an institution, sometimes centered around whether participants were aware of any plans to implement a certain measure, and if these measures were fully or partially implemented. Participants were also asked for their opinion on issues that were raised in the literature. To align the questionnaire with current library practices, a collaboration agreement was concluded with the DEAL Consortium, which contributed to the questionnaire development by sharing challenges that libraries currently face, commenting on and improving the questionnaire draft, and promoting the survey.

Discussions with staff members from the University Library of Humboldt-Universität zu Berlin who are responsible for recording publication costs yielded further insights into practical issues relating to data collection. In addition, a pretest was conducted with 11 respondents. Representatives of each institution type participated in the pretest, as did an expert from an RFO. Based on feedback from the pretest participants, the questionnaire was revised.

The final questionnaire comprised 26 questions: 16 single-choice questions, 4 multiple-choice questions, 2 matrix questions, and 4 open-ended questions. Seventeen questions offered respondents the option to provide additional comments via free-text fields. Ten questions were administered to all respondents, and 16 were administered conditional upon previous responses. Key terms were defined in infoboxes, with references to entries from a glossary that is maintained by a large German Open Access project, open-access.network.\footnote{Glossary of the open-access.network: \href{https://open-access.network/en/information/glossary}{https://open-access.network/en/information/glossary}.} The full questionnaire can be found in the Online Appendix. The questionnaire was implemented bilingually (in German and English) using the LimeSurvey instance of Humboldt-Universität zu Berlin. The data protection statement was coordinated with the university's data protection officer.

\subsection{Field Phase}
The survey was active from October 1 to November 12, 2024. During the field phase, two reminders were sent to invitees, on October 29 and November 5, 2024. Additionally, the survey was promoted via the mailing lists of the German Library Association (dbv) Sections 4 (academic universal libraries) and 5 (academic special libraries) and the mailing list of institutions participating in the DEAL agreements.

\subsection{Response Rate}
By the end of the field phase, 258 complete responses had been received, resulting in a response rate of 44.3\%. With a population of 583 and a sample size of 258, the maximum sampling error was 4.56\%. Table \ref{table:verteiler} shows the response rates for each institution type.

\begin{table}[!ht]
    \centering
    \footnotesize
    \begin{tabular}{p{5.5cm}p{2cm}p{2cm}p{2cm}}
    \hline
        Institution Type & Institutions in Initial List & Invited Institutions & Response Rate \\ \hline
        University of Applied Sciences & 186 & 186 & 39.8\% (74) \\ \hline
        Fraunhofer Society Institute & 76 & 76 & 6.6\% (5) \\ \hline
        Helmholtz Association Center & 18 & 17 & 58.8\% (10) \\ \hline
        Leibniz Association Institute & 96 & 95 & 48.4\% (46) \\ \hline
        Max Planck Institute & 91 & 77 & 40.3\% (31) \\ \hline
        Federal Departmental Research Institution & 45 & 44 & 50\% (22) \\ \hline
        University & 88 & 88 & 78.4\% (69) \\ \hline
    \end{tabular}
    \caption{Overview of contacted institutions and response rates}
    \label{table:verteiler}
\end{table}

\subsection{Data Preparation and Analysis of Open Ended Questions}
In preparation for the analysis, entries in open-text fields were reviewed and, where necessary, recoded. After examining the open-text fields, 103 comments were reassigned to an existing response option. Additionally, 71 responses to Question 16, which was open ended, were adjusted to yield numerical values where possible. 

The survey included two substantive open-ended questions (Questions 4 and 9). As the number of responses and the volume of text were too small for a traditional qualitative content analysis, the response texts were first summarized and then assigned descriptive codes. This approach allowed for a systematic description of the content. Code frequencies are not reported.

Results were analyzed and visualized using the statistical software R \autocite{r_core_team_r_2025} and functions from the tidyverse \autocite{tidyverse2019}.

\section{Results}
The survey results in this section are presented by question in the order in which the questions appeared in the questionnaire.
Responses to the final questions, Question 25 (``Do you have any additional remarks on collecting publication costs at your institution that you would like to share with us?'') and Question 26 (``Would you like to be informed when the results of this survey are published?''), are not reported here.

\subsection{Respondents' Areas of Work}
In response to Question 1, ``In which area of your institution do you work?'' (\emph{n} = 258), the vast majority (84.1\%, \emph{n} = 217) of respondents reported that they worked at the library, 6.6\% (\emph{n} = 17) stated that they worked in administration, and 6.2\% (\emph{n} = 16) that they were part of another central unit (e.g., research services or third-party funding office). Two respondents each (0.8\%) worked at the finance department or in research and teaching (e.g., at a faculty, a department, etc.); 1.6\% of respondents (\emph{n} = 4) did not answer this question.

\subsection{Monitoring Activities}
\begin{figure}
    \centering
    \includegraphics[width=0.9\linewidth]{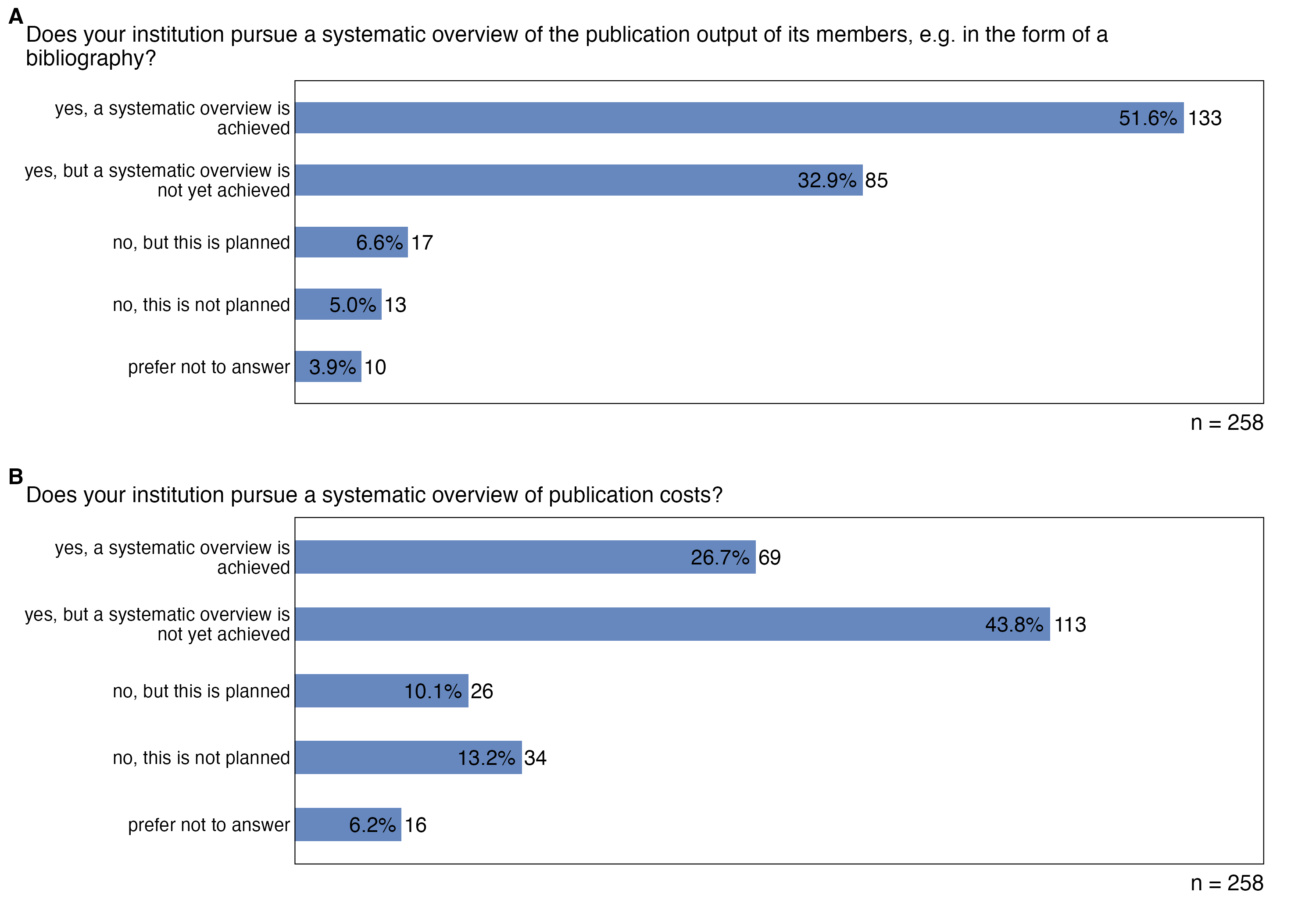}
    \caption{Systematic overview of publication output (A) and publication costs (B)}
    \label{fig:figure_1}
\end{figure}

Questions 2-4 addressed monitoring activities at the surveyed institutions. In response to Question 2, ``Does your institution pursue a systematic overview of the publication output of its members, e.g. in the form of a bibliography?'' (\emph{n} = 258; see Figure \ref{fig:figure_1} A), 51.6\% of respondents (\emph{n} = 133) reported that this had already been achieved; 32.9\% (\emph{n} = 85) reported that a systematic overview was planned but had not yet been achieved; 6.6\% (\emph{n} = 17) stated that their institutions did not systematically record publication output but planned to do so in the future; 5\% (\emph{n} = 13) reported that their institutions neither recorded nor planned to record such data; and 3.9\% (\emph{n} = 10) did not answer this question. Analysis of the results by institution type showed that a systematic overview of publication output had been achieved at 90\% of the Helmholtz Centers (\emph{n} = 9), 81.8\% of the federal departmental research institutions (\emph{n} = 18), and 78.3\% of the Leibniz Institutes (\emph{n} = 36) represented in the survey.

In response to Question 3, ``Does your institution pursue a systematic overview of publication costs?'' (\emph{n} = 258; see Figure \ref{fig:figure_1} B), 43.8\% of respondents (\emph{n} = 113) reported that an overview was planned but had not yet been achieved; 26.7\% (\emph{n} = 69) indicated that it was already in place; 13.2\% (\emph{n} = 34) stated that it was not planned; and 10.1\% (\emph{n} = 26) that it was planned. The remaining 6.2\% (\emph{n} = 16) did not answer this question. Analysis of the results by institution type revealed that a systematic overview of publication costs had been achieved at a minimum of 50\% of institutions of a specific type only at federal departmental research institutions (72.7\%, \emph{n} = 16) and Helmholtz Centers (50\%, \emph{n} = 5).

Question 4, ``Why does your institution not pursue a systematic overview of publication costs?'' (\emph{n} = 24), was an open-ended question administered only to respondents who selected the option ``No, this is not planned'' in Question 3. The most frequently cited reasons included a lack of necessity or resources at the institution. In some cases, processes or support from management were lacking, or the recording of publication costs was handled by a superordinate organization.

\subsection{Open Access Publication Funds}
Questions 5 to 10 of the questionnaire focused on publication funds. Responses to Question 5, ``Can publication costs at your institution be paid fully or partially by a publication fund?'' (\emph{n} = 258), showed that the vast majority (74\%, \emph{n} = 191) of the institutions represented in the survey had a publication fund in place; 14.3\% of respondents (\emph{n} = 37) stated that the implementation of a publication fund was not planned; 7\% (\emph{n} = 18) reported that it was planned; and 4.7\% (\emph{n} = 12) did not answer the question. An analysis of the results by institution type revealed that publication funds had been set up at 92.8\% of the universities (\emph{n} = 64) and 77.4\% of the Max Planck Institutes (\emph{n} = 24) represented in the survey.

Questions 6 - 8 were administered only if respondents answered ``Yes'' to Question 5 (``Can publication costs at your institution be paid fully or partially by a publication fund?''). In response to Question 6, ``Is partial funding of publication costs (cost splitting) possible at your institution?'' (\emph{n} = 184), 75\% of respondents (\emph{n} = 138) reported that cost splitting was possible; 13.6\% (\emph{n} = 25) reported that it was not planned at their institution; and 3.3\% (\emph{n} = 6) stated that it was planned. The remaining 8.2\% (\emph{n} = 15) did not answer this question.

Question 7 addressed the sources of funding for the publication fund (``Who provides the financial resources for the publication fund?'' \emph{n} = 184; multiple selection possible). By far the most frequently reported providers of financial resources were central service facilities (e.g., the library; 53.3\%, \emph{n} = 98) or the institution (47.8\%, \emph{n} = 88). Third-party funds acquired by central service facilities (26.1\%, \emph{n} = 48) or by the institution (22.8\%, \emph{n} = 42) were reported somewhat less frequently. Budget funds from research and teaching departments (14.7\%, \emph{n} = 27) and superordinate institutions (12\%, \emph{n} = 22) - for example, the Max Planck Digital Library (MPDL) in the case of the Max Planck Society - also contributed in some cases. Additionally, 8.2\% of respondents (\emph{n} = 15) selected the option ``third-party funds {[}acquired by{]} research and teaching''; 6.5\% (\emph{n} = 12) selected ``third-party funds {[}provided by a{]} federal state''; and 3.3\% (\emph{n} = 6) selected ``third-party funds {[}raised by a{]} superordinate institution. The remaining 1.1\% (\emph{n} = 2) did not answer the question. Respondents selected up to 6 options, with a median of 2 selections, suggesting that a publication fund at a typical institution was financed from two sources.

\begin{figure}
    \centering
    \includegraphics[width=0.9\linewidth]{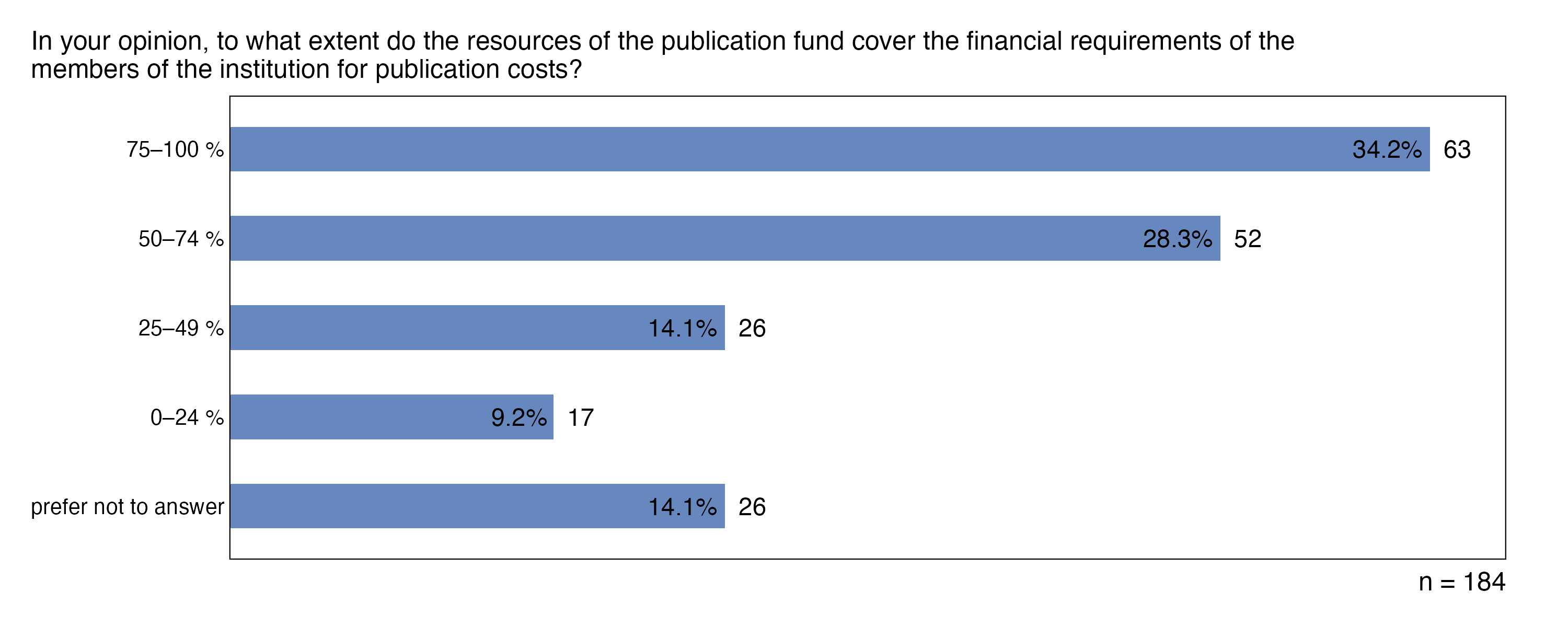}
    \caption{Coverage of financial needs by publication funds}
    \label{fig:figure_2}
\end{figure}

In response to Question 8, ``In your opinion, to what extent do the resources of the publication fund cover the financial requirements of the members of the institution for publication costs?'' (\emph{n} = 184), 34.2\% of respondents (\emph{n} = 63) stated that the publication fund covered ``75 - 100\%'' of members' financial requirements for publication costs (see Figure \ref{fig:figure_2}). A similar share of respondents (28.3\%, \emph{n} = 52) selected the option ``50 - 74\%''; 14.1\% (\emph{n} = 26) selected ``25 - 49\%''; and 9.2\% (\emph{n} = 17) selected ``0 - 24\%.'' The remaining 14.1\% (\emph{n} = 26) did not answer this question. An analysis of the results by institution type revealed that the publication fund covered 75 - 100\% of institution members' publication costs requirements at 86.7\% of the federal departmental research institutions (\emph{n} = 13). While not as high, above-average coverage of publication costs requirements was observed at 52.4 \% of the surveyed Max Planck Institutes (\emph{n} = 11), and 50\% of the surveyed Helmholtz Centers (\emph{n} = 3).

Questions 9 and 10 were administered only if respondents answered ``No, but this is planned'' or ``No, this is not planned'' to Question 5 (``Can publication costs at your institution be paid fully or partially by a publication fund?''). Responses to Question 9, ``What funds are used to pay publication costs at your institution?'' (\emph{n} = 46), which was open ended, revealed that institutions without a publication fund used similar funding sources to those with a fund. These sources largely corresponded to the response options provided in Question 7 (see Appendix). 

Among respondents from the 52 surveyed institutions without a publication fund, approximately one third agreed with the statement in Question 10, ``My institution considers researchers responsible for organizing the funding of publications.'' Specifically, 21.2\% of respondents (\emph{n} = 11) selected ``agree''; 13.5\% (\emph{n} = 7) selected ``strongly agree;'' 28.8\% (\emph{n} = 15) were ``undecided''; 19.2\% (\emph{n} = 10) disagreed strongly; and 7.7\% (\emph{n} = 4) disagreed. The remaining 9.6\% (\emph{n} = 5) did not answer this question.

\subsection{Agreements With Publishers}
Questions 11 and 12 of the questionnaire asked respondents about agreements with publishers. In response to Question 11 (``Does your institution participate in one or more of the DEAL contracts?''; \emph{n} = 258), 91.9\% of respondents (\emph{n} = 237) answered ``Yes.'' Of the respondents that answered ``No'' (5.4\%, \emph{n} = 14), none reported plans to participate in the future; 2.7\% (\emph{n} = 7) did not answer this question. 

In response to Question 12, ``Has your institution signed contracts with publishers besides the DEAL contracts that affect publication costs?'' (\emph{n} = 258), 76.4\% of respondents (\emph{n} = 197) answered ``Yes''; 16.3\% (\emph{n} = 42) answered ``No, this is not planned,'' and 1.9\% (\emph{n} = 5) answered ``No, but it is planned.'' The remaining 5.4\% (\emph{n} = 14) did not answer this question.

\subsection{Role of Management}
\begin{figure}
    \centering
    \includegraphics[width=0.9\linewidth]{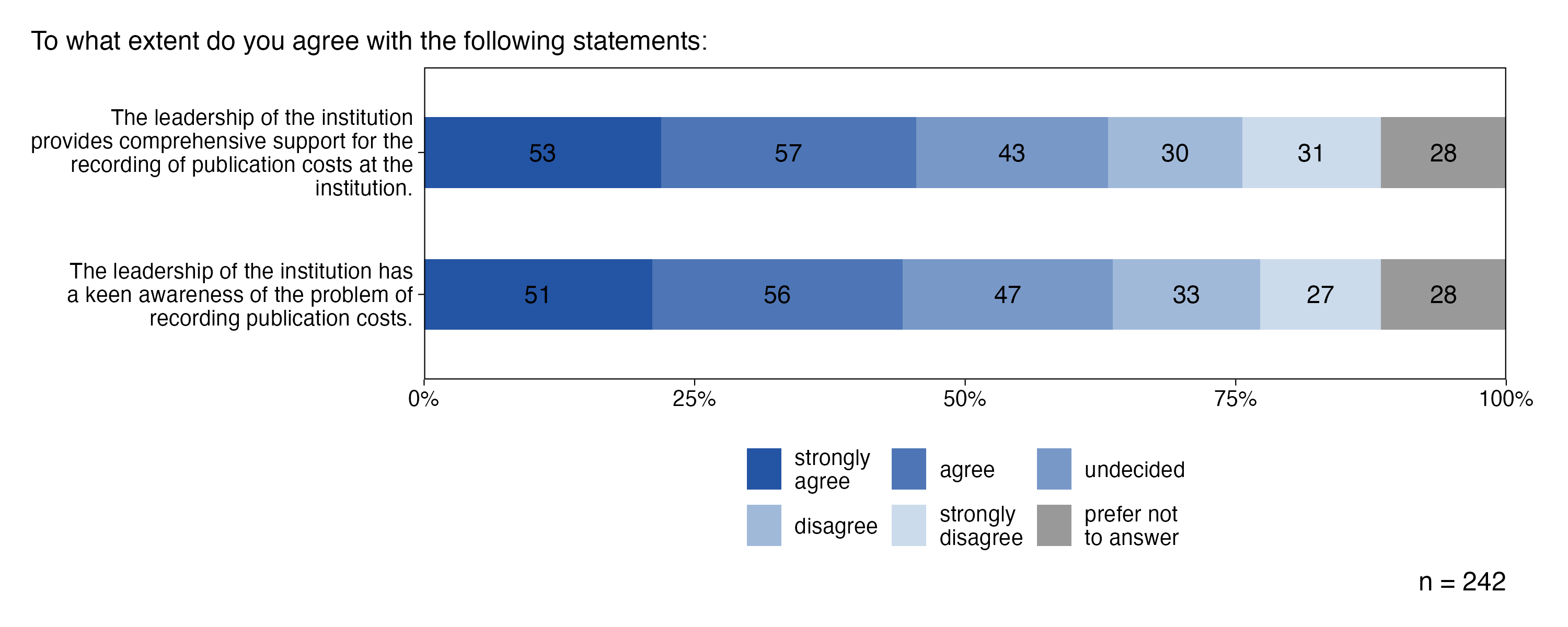}
    \caption{Assessment of (A) problem awareness and (B) support by the executive level}
    \label{fig:figure_3}
\end{figure}

Question 13 (\emph{n} = 242) was as a matrix question asking respondents to what extent they agreed with two statements regarding the attitude of management toward the recording of publication costs (see Figure \ref{fig:figure_3}). This question was administered only to respondents who did not identify as part of management (i.e., who did not select ``Administration'' in Question 1).

Asked to what extent they agreed with the statement in sub-question 13.1, ``The leadership of the institution has a keen awareness of the problem of recording publication costs,'' 23.1\% of respondents (\emph{n} = 56) selected ``agree,'' 21.1\% (\emph{n} = 51) ``strongly agree,'' 19.4\% (\emph{n} = 47) ``undecided,'' 13.6\% (\emph{n} = 33) ``disagree,'' and 11.2\% (\emph{n} = 27) ``strongly disagree.'' The remaining 11.6\% (\emph{n} = 28) did not answer the question. 

Asked about the extent of their agreement with the statement in sub-question 13.2, ``The leadership of the institution provides comprehensive support for the recording of publication costs at the institution,'' 23.6\% of respondents (\emph{n} = 57) selected ``agree,'' 21.9\% (\emph{n} = 53) ``strongly agree,'' 17.8\% (\emph{n} = 43) ``undecided,'' 12.8\% (\emph{n} = 31) ``disagree strongly,'' and 12.4\% (\emph{n} = 30) ``disagree.'' The remaining 11.6\% (\emph{n} = 28) did not respond.

\subsection{Workflows - Administrative Perspective}
Questions 14 to 16 addressed administrative aspects of designing workflows to record publication costs. These questions were administered only if respondents selected ``Yes, a systematic overview is achieved'' or ``Yes, but a systematic overview is not yet achieved'' in response to Question 3 (``Does your institution pursue a systematic overview of publication costs?'').

In response to Question 14, ``Are procedures for collecting publication costs clearly defined at your institution?'' (\emph{n} = 178), 42.7\% of respondents (\emph{n} = 76) answered: ``Yes, there are binding workflows, procedures or instructions''; 38.8\% (\emph{n} = 69) reported that workflows existed at their institution, but there were no binding specifications; 16.9\% (\emph{n} = 30) indicated that there were no workflows for recording publication costs at their institution; and 1.7\% (\emph{n} = 3) did not answer this question. Binding specifications for recording publication costs were in place at 88.2\% of surveyed federal departmental research institutions (\emph{n} = 15), at 52.5\% of universities (\emph{n} = 31), and 50\% of Helmholtz Centers (\emph{n} = 5). At all other surveyed institution types, binding specifications were in place at less than 50\% of the respective institutions.

\begin{figure}
    \centering
    \includegraphics[width=0.9\linewidth]{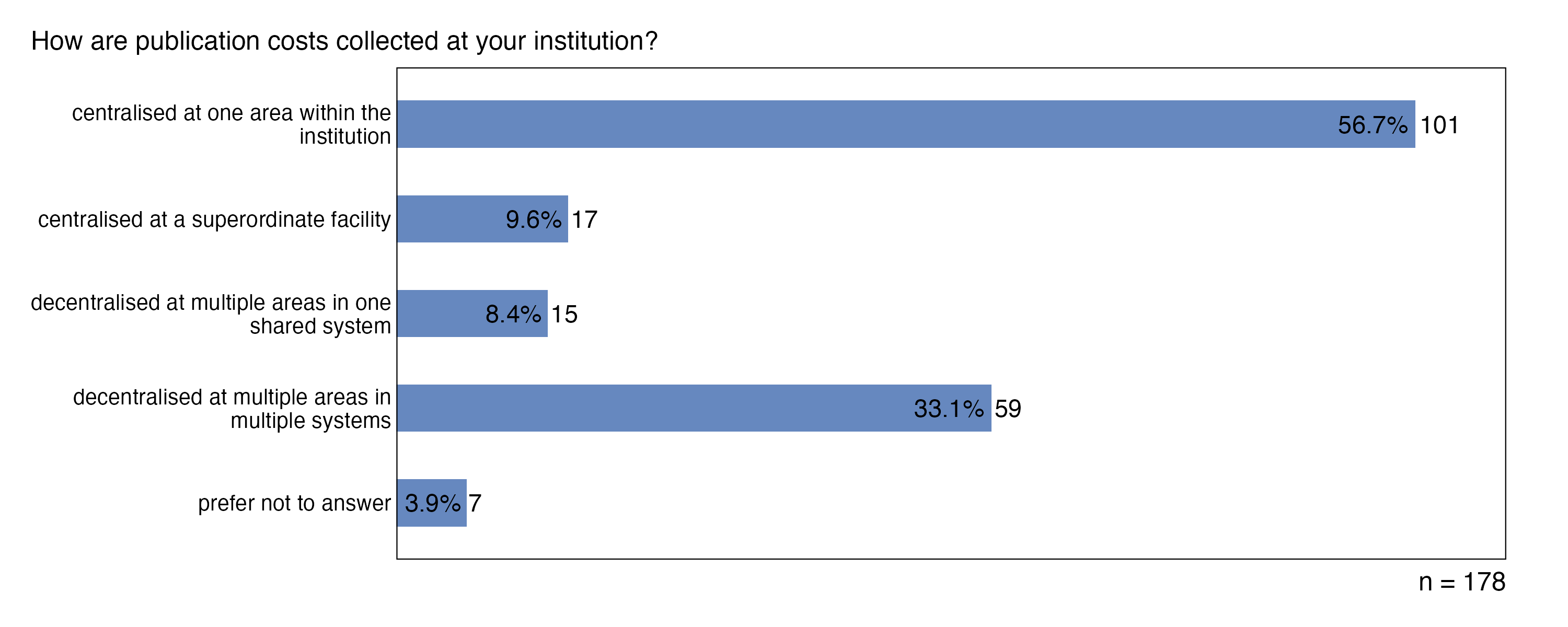}
    \caption{Recording of publication costs}
    \label{fig:figure_4}
\end{figure}

Responses to Question 15 (``How are publication costs collected at your institution?'' \emph{n} = 178; multiple selection possible; see Figure \ref{fig:figure_4}) showed that publication costs were recorded centrally by one department at 56.7\% of the surveyed institutions (\emph{n} = 101) and decentrally across multiple departments and systems at 33.1\% (\emph{n} = 59) of the surveyed institutions. A superordinate institution---for example, the Max Planck Digital Library (MDPL) in the case of the Max Planck Society---recorded publication costs centrally for 9.6\% (\emph{n} = 17) of the surveyed institutions. Additionally, 8.4\% of respondents (\emph{n} = 15) reported that costs were recorded decentrally across multiple departments but within a shared system. The remaining 3.9\% (\emph{n} = 7) did not answer this question. Participants selected up to 3 options, with a median of 1 selection. Publication costs were recorded centrally at one area within the institution at 87.5\% of surveyed federal departmental research institutions (\emph{n} = 14) and were recorded centrally at a superordinate facility at 77.8\% of surveyed Max Planck Institutes (\emph{n} = 14). At all other surveyed institution types, central recording of publication costs was below 75\%.

Question 16, ``How many publications by members of your institution do you estimate have incurred publication costs charged to your institution in 2023?'' (\emph{n} = 123), was an open-ended question. Respondents provided estimates ranging from 1 to 3,000 publications, with a median of 35 and an average of 226.4. At 407.6, the standard deviation was very high.

\subsection{Workflows - Practice}
Questions 17 to 21 on the specific design of workflows for recording publication costs. These questions were administered only to those respondents who selected the option ``Yes, a systematic overview is achieved'' or ``Yes, but a systematic overview is not yet achieved'' in response to Question 3 (``Does your institution pursue a systematic overview of publication costs?'').

Responses to Question 17, ``Which areas of your institution are involved in collecting publication costs?'' (\emph{n} = 178; multiple selection possible), show that the library (87.6\%, \emph{n} = 156) and the finance department (78.1\%, n = 139) were involved in recording publication costs at most of the surveyed RPOs. Another central department (e.g., research services or third-party funding office) was involved at 38.2\% (\emph{n} = 68) of the institutions. Departments from the areas of research and teaching were involved at 25.3\% of the institutions (\emph{n} = 45), and management was involved at 9\% of the institutions (\emph{n} = 16). The remaining 1.1\% of respondents (\emph{n} = 2) did not answer this question. Participants selected up to 4 options, with a median of 2. This suggests that at least two departments at each institution were involved in recording publication costs.

Responding to Question 18, ``How do you evaluate the cooperation of the areas involved regarding the collection of publication costs?'' (\emph{n} = 178), 43.3\% of respondents (\emph{n} = 77) rated it as ``rather good,'' 23\% (\emph{n} = 41) as ``very good,'' 19.1\% (\emph{n} = 34) as ``average,'' and 5.1\% (\emph{n} = 9) as ``rather poor.'' None of the respondents selected the option ``very poor.''; 9.6\% of respondents (\emph{n} = 17) did not answer this question.

In response to Question 19, ``Which tools are used at your institution to collect publication costs?'' (\emph{n} = 178; multiple selection possible), most respondents reported that their institutions used a spreadsheet program (74.2\%, \emph{n} = 132) or a financial management system (66.3\%, \emph{n} = 118) for this purpose. The use of dashboards, for example, those provided by publishers, was reported by 48.3\% of respondents (\emph{n} = 86); library management systems were reported by 33.7\% (\emph{n} = 60); and repositories or comparable collections of publications or publication data by 24.7\% (\emph{n} = 44). Further, 17.4\% of respondents (\emph{n} = 31) reported that a customized database had been created at their institution. Less commonly used tools included ticket systems (9.6\%, \emph{n} = 17), current research information systems (8.4\%, \emph{n} = 15), and wikis (2.8\%, \emph{n} = 5); 3.9\% of respondents (\emph{n} = 7) did not answer this question. Respondents selected up to eight options, with a median of 3 selections, suggesting that approximately three different tools were used at each institution to record publication costs.

\begin{figure}
    \centering
    \includegraphics[width=0.9\linewidth]{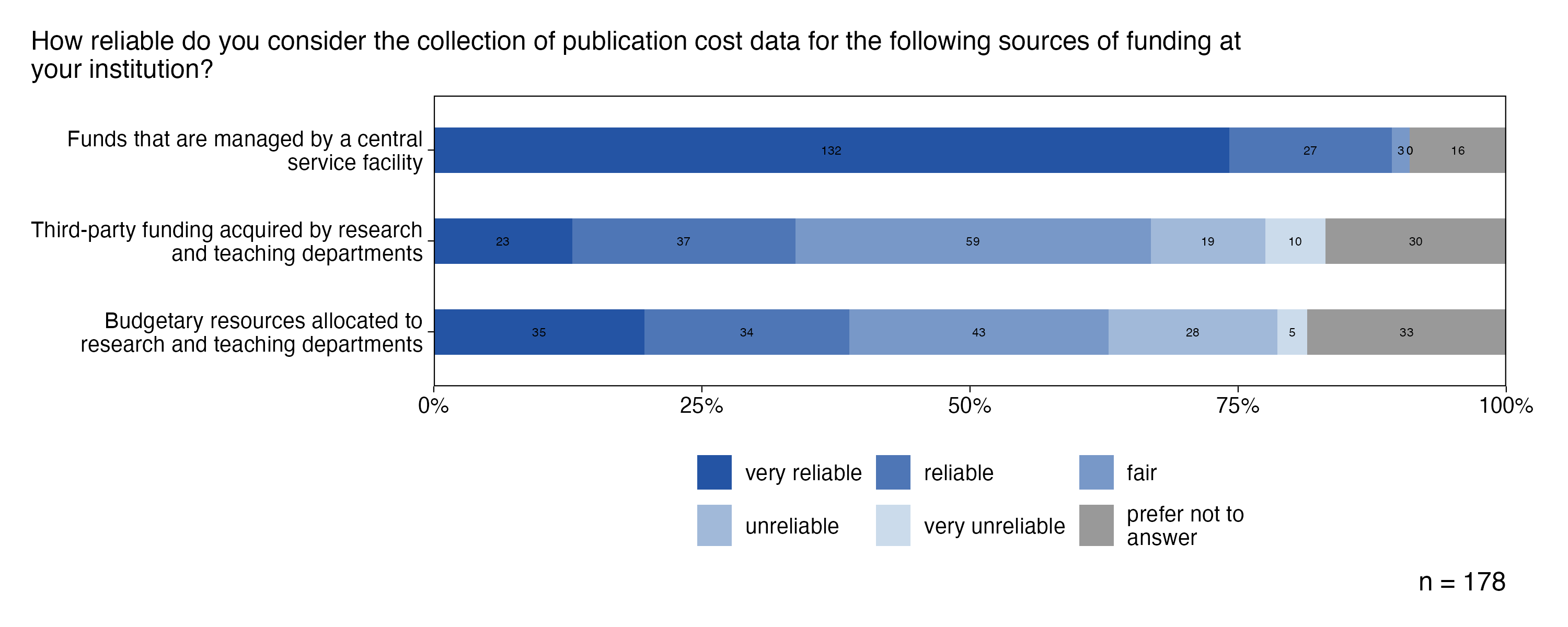}
    \caption{Assessment of the reliability of recording (A) funds managed by a central service unit; (B) third-party funds raised by research and teaching; and (C) institutional budget funds allocated to research and teaching}
    \label{fig:figure_5}
\end{figure}

Question 20, ``How reliable do you consider the collection of publication cost data for the following sources of funding at your institution?'' (\emph{n} = 178), was a matrix question asking respondents to evaluate the reliability of data collection for specific types of funds (see Figure \ref{fig:figure_5}). Responding to sub-question 20.1, ``funds that are managed by a central service facility (e.g. via a publication fund),'' 74.2\% of respondents (\emph{n} = 132) rated the reliability of the collection of data on publication costs funded from this source as ``very reliable,'' 15.2\% (\emph{n} = 27) rated it as ``reliable,'' and 1.7\% (\emph{n} = 3) as ``fair.'' None of the respondents selected the options ``unreliable'' or ``very unreliable''; 9\% (\emph{n} = 16) did not answer this sub-question.

Responding to sub-question 20.2, ``third-party funding acquired by research and teaching departments (e. g. faculty, institute, department, section),'' 33.1\% of respondents (\emph{n} = 59) rated the reliability of the collection of data on publication costs funded from this source as ``fair,'' 20.8\% (\emph{n} = 37) as ``reliable,'' 12.9\% (\emph{n} = 23) as ``very reliable,'' 10.7\% (\emph{n} = 19) as ``unreliable,'' and 5.6\% (\emph{n} = 10) as ``very unreliable.'' The remaining 16.9\% (\emph{n} = 30) did not answer this sub-question.

In response to sub-question 20.3, ``budgetary resources allocated to research and teaching departments (e. g. faculty, institute, department, section),'' 24.2\% of respondents (\emph{n} = 43) rated the reliability of the collection of data on publication costs funded from this source as ``fair,'' 19.7\% (\emph{n} = 35) as ``very reliable,'' 19.1\% (\emph{n} = 34) as ``reliable'' 15.7\% (\emph{n} = 28) as ``unreliable,'' and 2.8\% (\emph{n} = 5) as ``very unreliable.'' The remaining 18.5\% (\emph{n} = 33) did not answer this sub-question.

\begin{figure}
    \centering
    \includegraphics[width=0.9\linewidth]{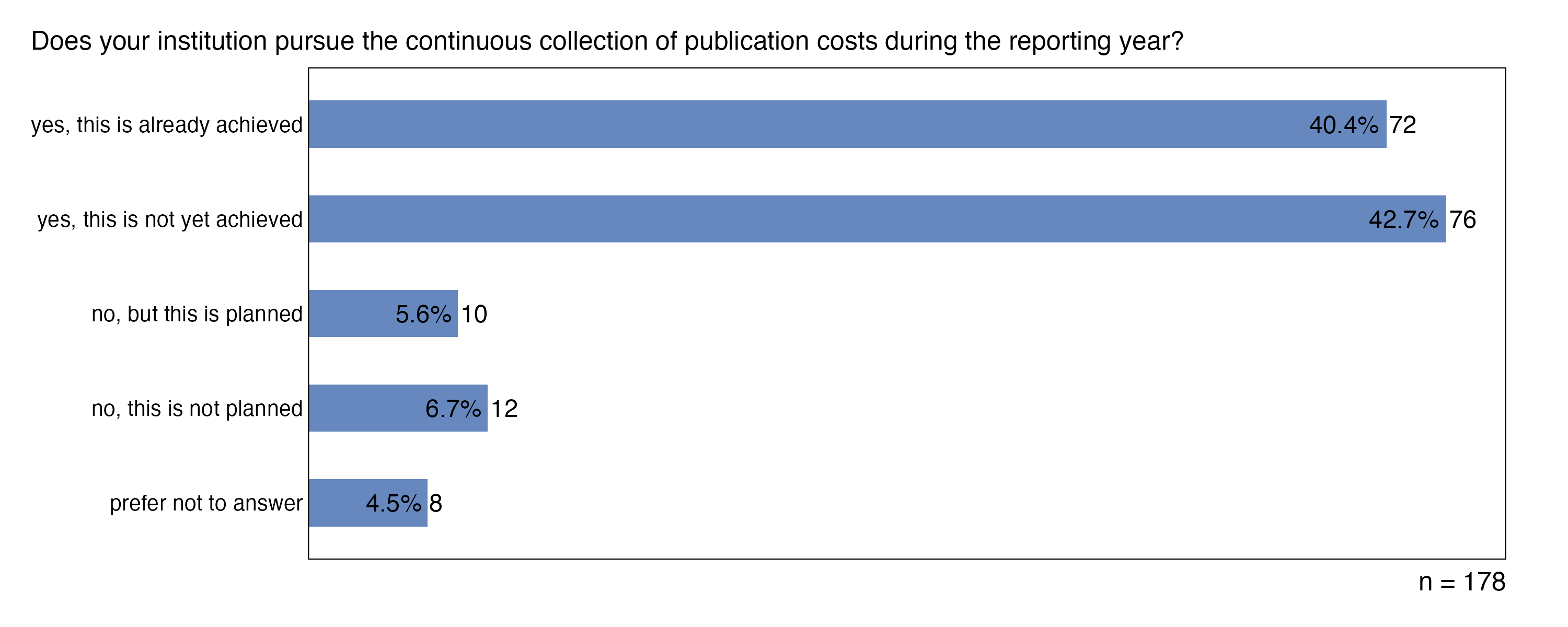}
    \caption{Continuous recording of publication costs during a reporting year}
    \label{fig:figure_6}
\end{figure}

Responding to Question 21, ``Does your institution pursue the continuous collection of publication costs during the reporting year?" (\emph{n} = 178), 42.7\% of respondents (\emph{n} = 76) stated that it was pursued but had not yet been achieved (see Figure \ref{fig:figure_6}); 40.4\% (\emph{n} = 72) reported that this goal had already been achieved; 6.7\% (\emph{n} = 12) stated that it was not planned; and 5.6\% (\emph{n} = 10) indicated that it was planned. The remaining 4.5\% (\emph{n} = 8) did not answer this question.

\subsection{Contribution to the Open Access Transformation}
Questions 22 to 24 of the questionnaire addressed the perceived contribution of recording publication costs to the open access transformation overall. Question 22 was a filter question and was administered only if respondent selected ``Yes, a systematic overview is achieved'' or ``Yes, but a systematic overview is not yet achieved'' in Question 3.

In response to Question 22, ``Are the collected publication costs used for strategic decisions at your institution (e.g. as part of resource planning)?'' (\emph{n} = 178), 59\% of respondents (\emph{n} = 105) answered ``Yes,'' and 25.3\% (\emph{n} = 45) answered ``No.'' The remaining 15.7\% (\emph{n} = 28) did not answer this question.

\begin{figure}
    \centering
    \includegraphics[width=0.9\linewidth]{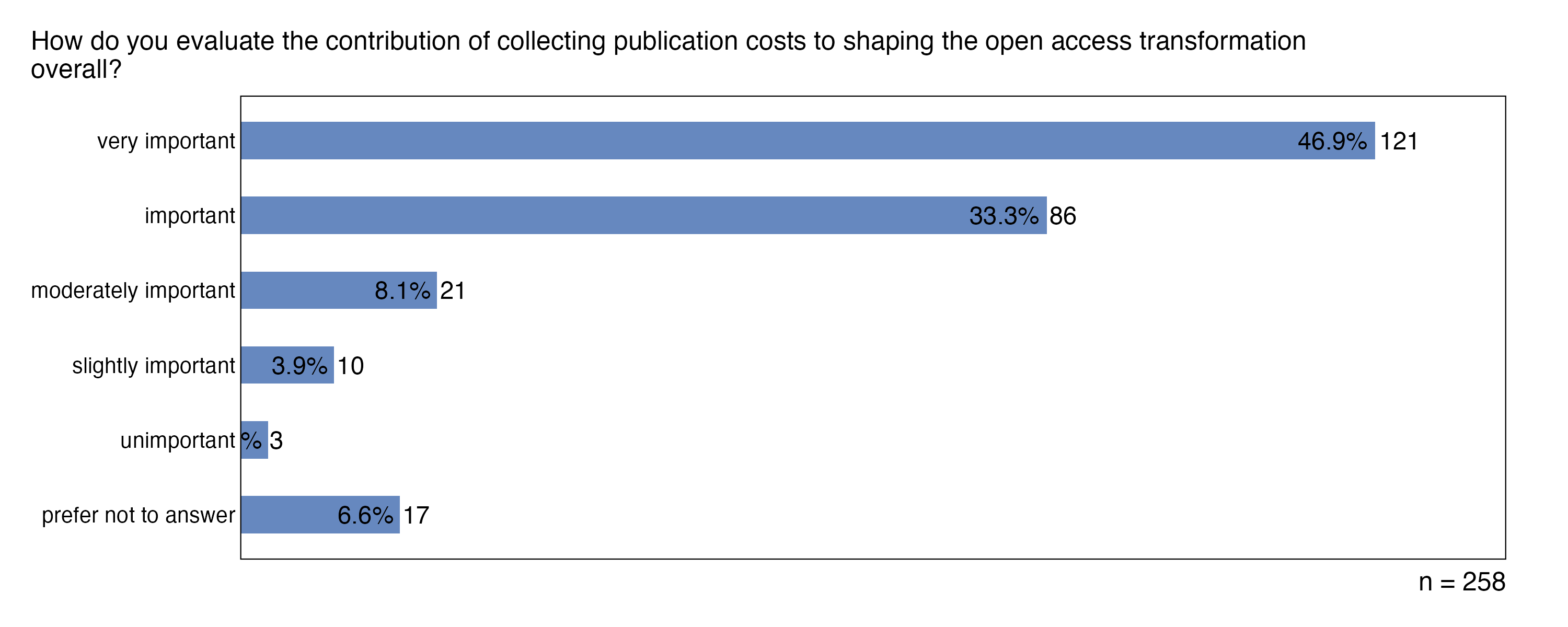}
    \caption{Assessment of the contribution of recording publication costs to the design of the Open Access transition}
    \label{fig:figure_7}
\end{figure}

Responding to Question 23, ``How do you evaluate the contribution of collecting publication costs to shaping the open access transformation overall?'' (\emph{n} = 258; see Figure \ref{fig:figure_7}), 46.9\% of respondents (\emph{n} = 121) rated it as ``very important,'' 33.3\% (\emph{n} = 86) as ``important,'' 8.1\% (\emph{n} = 21) as ``moderately important,'' 3.9\% (\emph{n} = 10) as ``slightly important,'', and 1.2\% (\emph{n} = 3) as ``unimportant.''. The remaining 6.6\% (\emph{n} = 17) did not answer this question. The contribution of recording publication costs to shaping the open access transformation was considered "very important" by 80\% of respondents from Helmholtz Centers (\emph{n} = 8), 53.6\% of respondents from universities (\emph{n} = 37), and 51.4\% of respondents from universities of applied sciences (\emph{n} = 38).

\begin{figure}
    \centering
    \includegraphics[width=0.9\linewidth]{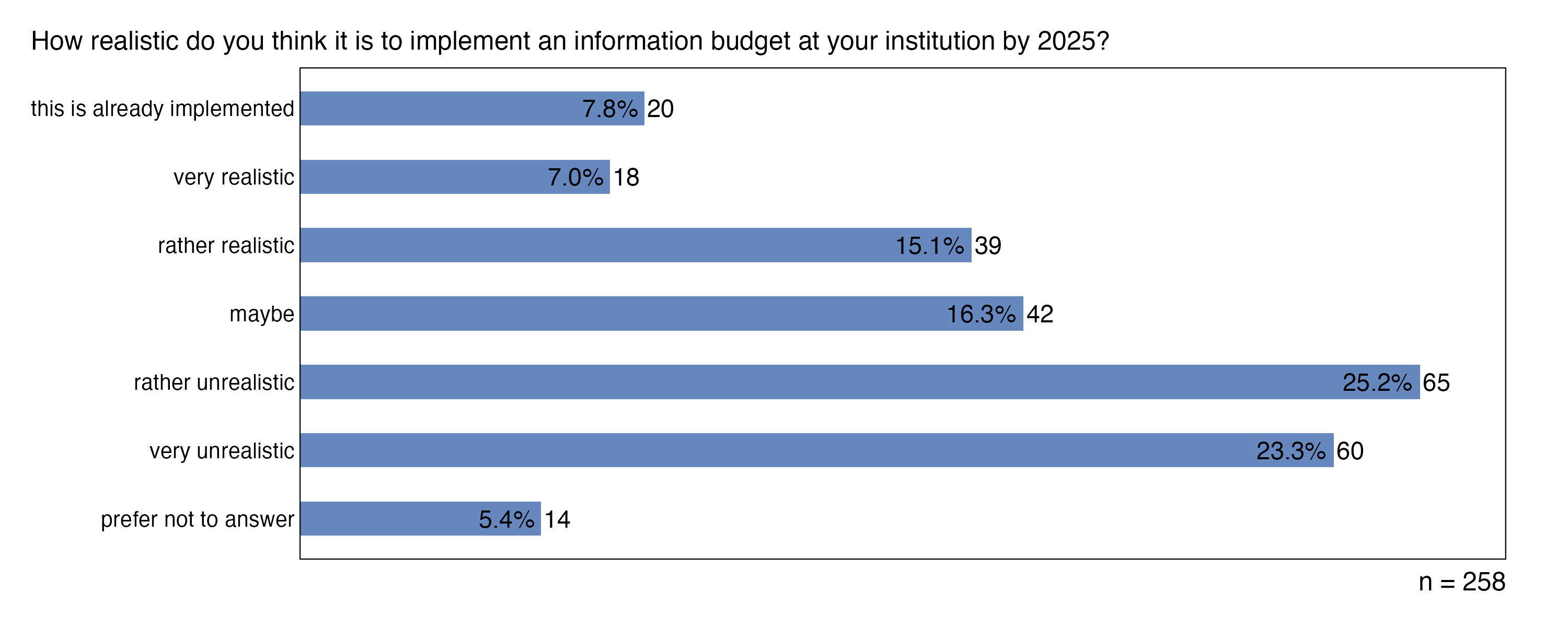}
    \caption{Assessment of the implementation of an information budget by 2025}
    \label{fig:figure_8}
\end{figure}

In response to Question 24, ``How realistic do you think it is to implement an information budget at your institution by 2025?'' (\emph{n} = 258; see Figure \ref{fig:figure_8}), 25.2\% (\emph{n} = 65) of respondents selected ``rather unrealistic,'' 23.3\% (\emph{n} = 60) ``very unrealistic,'' 16.3\% (\emph{n} = 42) "maybe", 15.1\% (\emph{n} = 39) "rather realistic,'' 7.8\% (\emph{n} = 20) ``already implemented,'' and 7\% (\emph{n} = 18) ``very realistic.'' The remaining 5.4\% (\emph{n} = 14) did not respond.

\section{Discussion}
The fragmentation and obfuscation of financial flows in the context of academic publishing outlined in the Introduction poses a challenge to RPOs. Library budgets are no longer the only source of funding, and achieving a comprehensive overview and transparency of all publication-related income streams and expenditures requires new workflows and additional resources. The survey presented in this article was the first comprehensive study on the collection of data on publication costs at RPOs in Germany, where the topic has been discussed for several years now under the label ``information budget''.

\subsection{Establishing Monitoring Activities}
The survey results indicate that monitoring activities---both for publication output and publication costs--- are not yet universally established at German RPOs. Only about half of the respondents reported that their institutions had achieved a systematic overview of publication output, and just 26.7\% reported that their institutions had achieved a systematic overview of publication costs. These monitoring activities are closely interlinked, as determining an institution's publication output is a prerequisite for the reliable and comprehensive recording of publication costs. Therefore, both activities could be integrated at RPOs. Some institutions already do this. For example, they use their library management system or institutional repository to simultaneously add information on publication costs to a database of publications they already maintain.

\subsection{Financing the Open Access Transformation}
Open access publication funds at German RPOs have been supported by the DFG, Germany's largest RFO, since 2010 \parencite{fournier2013} and are now widely used, as evidenced by the fact that 74\% of the institutions represented in the survey had a publication fund in place. Libraries contribute significantly to financing these publication funds, both by allocating budget funds and acquiring third-party funding. This highlights the central role that libraries play in the financing of open access publications. At 62.5\% of the surveyed RPOs with a publication fund, the fund covered between 50\% and 100\% of members' financial requirements for publication costs. However, a sizable funding gap was apparent at the circa 23\% of institutions where between 0\% and 49\% of financial requirements were covered. Pooling resources in publication funds seems to positively impact the reliability of recording publication costs, as publication costs covered by a publication fund are recorded more reliably compared with other sources of funding.

\subsection{Fully Realizing the Potential Inherent in Recording Publication Costs}
Eighty percent of respondents considered the recording of publication costs to be ``very important'' or ``important'' for the open access transformation. However, our survey results also revealed that they saw room for improvement in terms of management's problem awareness and support for the collection of data on publication costs. Moreover, the fact that only 59\% of surveyed institutions used the collected data in strategic decision-making suggests that the potential inherent in the recording of publication costs has not yet been fully realized.

RPOs pursue diverse approaches to recording publication costs. At many RPOs, these procedures could be further standardized in the future. Although procedures were often implemented at the surveyed institutions, they were not formally specified. It is notable that data collection tended to occur either centrally within one department or decentrally across several departments and systems. Data collected centrally, for example, by administrators of publication funds, were perceived as more reliable, whereas decentralized approaches were perceived as more prone to errors.

Recording publication costs is a collaborative task involving several departments within an institution, typically the library and the finance department. Although some respondents felt that interdepartmental collaboration could be improved at their institutions, it was generally rated as good. The survey results also indicate that spreadsheet programs are often used for recording publication costs. This is due in part to the fact that the DFG historically required RPOs to use an Excel spreadsheet to report funds received under its program ``Open Access Publication Funding.'' However, it also indicates that a significant amount of manual work is involved in the collection of data on publication costs, and that data exchange between systems is lacking. In summary, the results point to a clear need to further support workflows and develop tools for aggregating data on publication costs.

\subsection{Developing Information Budgets}
Although the concept of the information budget has not (yet) established itself internationally, it could prove useful in cases where libraries want to obtain an overview of all income streams and expenditures relating to scientific information. When discussing the financial dimension of open access, the fact that the subscription model still dominates the for-profit publishing market is sometimes overlooked \parencite{pollock_news_2025}. Collating all scientific-information-related income streams and expenditures of an RPO in an information budget can help to relate them to each other.
 
Only a small percentage of respondents reported that their RPOs had already established an information budget. Despite the German Science and Humanities Council (2022) recommendation that RPOs should implement an information budget by 2025, the concept remains uncommon in Germany, and a large share of respondents considered the implementation of an information budget at their institution by 2025 unlikely. However, the survey also shows that some institutions have already established the necessary groundwork for an information budget. For instance, several institutions represented in the survey were monitoring publication output or costs or were working toward these goals. Many had established or were planning to implement continuous recording of publication costs during the fiscal year, which is another prerequisite for an information budget. Other factors---in particular the integration of data from different sources---were still lacking, which hinders the implementation of an information budget.

The survey results highlight the central role that libraries play in financing the open access transformation: At most institutions with publication funds, libraries were contributing to their financing. However, publication funds covered 75--100\% of members' financial requirements for publication costs at only around one third of the surveyed institutions. At some institutions represented in the survey, covering publication costs was still seen as the responsibility of individual authors, which could pose a challenge to achieving full cost transparency at RPOs.

\subsection{A Global Initiative for Cost Transparency}
Although the present survey focused solely on Germany, RPOs in other regions are facing similar challenges and are taking steps to establish workflows \parencite{collister_preparing_2025,kemp_cost_2024,, neylon_trajectory_2026}. As academic publishing is an international market, achieving cost transparency requires a global initiative. Initial steps in this direction have already been taken, as the initiatives described in the Introduction demonstrate. However, there are still considerable barriers to achieving full cost transparency. For example, subscription agreements with publishers often include non-disclosure clauses. In recent years, several RPOs and other stakeholders have published negotiating principles indicating that they will no longer accept this \parencite{esac_initiative_negotiation_nodate}. And although RPOs and RFOs from many countries are participating in the data collection initiative OpenAPC, there are still gaps in geographic coverage \parencite[6297]{asai2023}.

\subsection{Limitations}
The present study has several limitations. Given the diversity of the German research system, the survey focused on a wide range of institution types, and the response rates varied across institution types (see Table \ref{table:verteiler}). At 6.6\%, the response rate for Fraunhofer Society Institutes was particularly low, and the survey results therefore offer limited insights for this institution type. Additionally, the order of questions within a questionnaire can influence responses. For instance, it is possible that responses to Question 23 (``How do you evaluate the contribution of collecting publication costs to shaping the open access transformation overall?'') were more positive because it appeared toward the end of the survey.

\section{Conclusion}
The results of the present survey indicate that data on publication costs could be collected more precisely at German RPOs. Although the overwhelming majority of respondents recognized that recording publication costs was important for the open access transformation, data collection efforts at the surveyed institutions were complicated by a lack of mandatory workflows, institutional support, and interdepartmental collaboration.

RPOs could benefit from transforming existing instruments, such as open access publication funds, into information budgets. This would create a systematic overview of all information-related income streams and expenditures and could provide a basis for decision-making. A clear benefit would be that RPOs could consider various funding and business models and act more in line with their goals for transforming academic publishing.

In recent years, the dependency on commercial services for research information has been increasingly criticized. The signatories of the Barcelona Declaration on Open Research Information,\footnote{Barcelona Declaration on Open Research Information: \url{https://barcelona-declaration.org}} among them many RPOs, commit to the openness of research information. More than 750 institutions have contributed to OpenAPC, an initiative for collecting cost data.\footnote{OpenAPC - Institutions: \url{https://github.com/OpenAPC/openapc-de/blob/master/data/institutions.csv}; accessed: 2026-07-14} Collecting and publishing data on their publication-related expenditures adds valuable insights into the financial dimension of academic publishing to the growing open research information ecosystem. \citeauthor{neylon_trajectory_2026} stress that "[T]here is substantial value in making this financial data more public and transparent, as it can support comparison, discussion and critique. [Where] data is shared it helps us to build a picture of global investments and plan as a community for the future." \parencite[40]{neylon_trajectory_2026}
\\
The path to global cost transparency is challenging, but worthwhile.

\section*{Author Contribution Statement}
\begin{itemize}
    \item[] Dorothea Strecker: Data Curation ; Formal Analysis ; Investigation ; Methodology ; Project Administration ; Visualization ; Writing – Original Draft ; Writing – Review \& Editing
    \item[] Heinz Pampel: Conceptualization ; Funding Acquisition ; Methodology ; Project Administration ; Resources ; Supervision ; Writing – Review \& Editing
    \item[] Jonas Höfting: Data Curation ; Formal Analysis ; Investigation ; Methodology ; Visualization ; Writing – Review \& Editing
\end{itemize}

\section*{Acknowledgements}
We would like to thank Ida Mäder, Tanya Lackner, and Melanie Schreiber from the University Library of Humboldt-Universität zu Berlin, as well as the pretest participants, for their support in developing the questionnaire. Furthermore, we express our gratitude to Karin Ilg, Jochen Johannsen, Oliver Kohl-Frey, and Katrin Stump from the DEAL group for contributing to the development of the questionnaire.

\section*{Funding Information}
The survey was conducted as part of the project "Datenpraxis zur Gestaltung der Open-Access-Transformation - Analyse, Empfehlung, Training \& Vernetzung (OA Datenpraxis)", funded by the German Research Foundation (DFG) (project number 528466070). Heinz Pampel’s work was supported by the Einstein Center Digital Future (ECDF).

\section*{Data Availability}
The data underlying the analysis has been anonymized and made available via Zenodo \autocite{strecker2025}:

Strecker, D., Pampel, H., \& Höfting, J. (2025). Dataset for: Recording of publication costs at research performing institutions in Germany (Version 1.0) [Dataset]. Zenodo. \url{https://doi.org/10.5281/zenodo.14732554}

\newpage

\section*{Appendix A: Questionnaire} \label{sec:questionnaire}  
\footnotesize
\begin{longtable}{
>{\raggedright\arraybackslash}p{0.3cm}
>{\raggedright\arraybackslash}p{3.7cm}
>{\raggedright\arraybackslash}p{5cm}
>{\raggedright\arraybackslash}p{1.3cm}
>{\raggedright\arraybackslash}p{2.7cm}}

\hline
No. & Question text & Answer options & Type & Condition \\ \hline
\endfirsthead

\hline
No. & Question text & Answer options & Type & Condition \\ \hline
\endhead

\hline
\multicolumn{5}{c}{\small Continued on next page} \\
\endfoot

\hline
\endlastfoot
        01 & In which area of your institution do you work? & 
        central services: library \linebreak
        central services: other (e.g. research services) \linebreak
        administration \linebreak
        finance department \linebreak
        research and teaching (e.g. faculty, institute, department, section, individual researchers) \linebreak
        prefer not to answer \linebreak
        other & single choice & none \\ \hline

        02 & Does your institution pursue a systematic overview of the publication output of its members, e.g. in the form of a bibliography? & 
        yes, a systematic overview is achieved \linebreak
        yes, but a systematic overview is not yet achieved \linebreak
        no, but this is planned \linebreak
        no, this is not planned \linebreak
        prefer not to answer \linebreak
        other & single choice & none \\ \hline

        03 & Does your institution pursue a systematic overview of publication costs? & 
        yes, a systematic overview is achieved \linebreak
        yes, but a systematic overview is not yet achieved \linebreak
        no, but this is planned \linebreak
        no, this is not planned \linebreak
        prefer not to answer \linebreak
        other & single choice & none \\ \hline

        04 & Why does your institution not pursue a systematic overview of publication costs? & 
        NA & free text & answered “no, this is not planned” in Question 3 \\ \hline

        05 & Can publication costs at your institution be paid fully or partially by a publication fund? & 
        yes \linebreak
        no, but this is planned \linebreak
        no, this is not planned \linebreak
        prefer not to answer \linebreak
        other & single choice & none \\ \hline

        06 & Is partial funding of publication costs (cost splitting) possible at your institution? & 
        yes \linebreak
        no, but this is planned \linebreak
        no, this is not planned \linebreak
        prefer not to answer \linebreak
        other & single choice & answered “yes” in Question 5 \\ \hline

        07 & Who provides the financial resources for the publication fund? & 
        institution: budget funds \linebreak
        institution: third-party funds \linebreak
        central service facility (e.g. library) at the institution: budget funds \linebreak
        central service facility (e.g. library) at the institution: third-party funds \linebreak
        superordinate institution (e.g. the MPDL at the MPG): budget funds \linebreak
        superordinate institution (e.g. the MPDL at the MPG): third-party funds \linebreak
        research and teaching (e.g. faculty, institute, department, chair, section, individual researchers): budget funds \linebreak
        research and teaching (e.g. faculty, institute, department, chair, section, individual researchers): third-party funds \linebreak
        federal state \linebreak
        prefer not to answer \linebreak
        other & Multiple choice & answered “yes” in Question 5 \\ \hline

        08 & In your opinion, to what extent do the resources of the publication fund cover the financial requirements of the members of the institution for publication costs? & 
        0–24\% \linebreak
        25–49\% \linebreak
        50–74\% \linebreak
        75–100\% \linebreak
        prefer not to answer \linebreak
        other & single choice & answered “yes” in Question 5 \\ \hline

        09 & What funds are used to pay publication costs at your institution? & 
        NA & free text & answered “no, but this is planned” or “no, this is not planned” in Question 5 \\ \hline

        10 & To what extent do you agree with the following statement: My institution considers researchers responsible for organizing the funding of publications. & 
        strongly agree \linebreak
        agree \linebreak
        undecided \linebreak
        disagree \linebreak
        strongly disagree \linebreak
        prefer not to answer \linebreak
        other & single choice & answered “no, but this is planned” or “no, this is not planned” in Question 5 \\ \hline

        11 & Does your institution participate in one or more of the DEAL contracts? & 
        yes \linebreak
        no, but this is planned \linebreak
        no, this is not planned \linebreak
        prefer not to answer \linebreak
        other & single choice & none \\ \hline

        12 & Has your institution signed contracts with publishers besides the DEAL contracts that affect publication costs? & 
        yes \linebreak
        no, but this is planned \linebreak
        no, this is not planned \linebreak
        prefer not to answer \linebreak
        other & single choice & none \\ \hline

        13 & To what extent do you agree with the following statements: \linebreak
        13.1: The leadership of the institution has a keen awareness of the problem of recording publication costs. \linebreak
        13.2: The leadership of the institution provides comprehensive support for the recording of publication costs at the institution. & 
        strongly agree \linebreak
        agree \linebreak
        undecided \linebreak
        disagree \linebreak
        strongly disagree \linebreak
        prefer not to answer & matrix & did not answer “administration” in Question 1 \\ \hline

        14 & Are procedures for collecting publication costs clearly defined at your institution? & 
        yes, there are binding workflows, procedures or instructions \linebreak
        yes, but there are no binding specifications \linebreak
        no \linebreak
        prefer not to answer \linebreak
        other & single choice & answered “yes, a systematic overview is achieved” or “yes, but a systematic overview is not yet achieved” in Question 3 \\ \hline

        15 & How are publication costs collected at your institution? & 
        centralised at one area within the institution \linebreak
        centralised at a superordinate facility (e.g. the MPDL at the MPG) \linebreak
        decentralised at multiple areas in one shared system \linebreak
        decentralised at multiple areas in multiple systems \linebreak
        prefer not to answer \linebreak
        other & Multiple choice & answered “yes, a systematic overview is achieved” or “yes, but a systematic overview is not yet achieved” in Question 3 \\ \hline

        16 & How many publications by members of your institution do you estimate have incurred publication costs charged to your institution in 2023? & 
        NA & free text & answered “yes, a systematic overview is achieved” or “yes, but a systematic overview is not yet achieved” in Question 3 \\ \hline

        17 & Which areas of your institution are involved in collecting publication costs? & 
        central services: library \linebreak
        central services: other (e.g. research services) \linebreak
        administration \linebreak
        finance department \linebreak
        research and teaching (e.g. faculty, institute, department, section, individual researchers) \linebreak
        prefer not to answer \linebreak
        other & Multiple choice & answered “yes, a systematic overview is achieved” or “yes, but a systematic overview is not yet achieved” in Question 3 \\ \hline

        18 & How do you evaluate the cooperation of the areas involved regarding the collection of publication costs? & 
        very good \linebreak
        good \linebreak
        fair \linebreak
        poor \linebreak
        very poor \linebreak
        prefer not to answer & single choice & answered “yes, a systematic overview is achieved” or “yes, but a systematic overview is not yet achieved” in Question 3 \\ \hline

        19 & Which tools are used at your institution to collect publication costs? & 
        financial management system \linebreak
        library management system \linebreak
        research information system \linebreak
        dashboard (e.g. publisher dashboards) \linebreak
        spreadsheet software \linebreak
        repository, publication server, publication database, institutional bibliography \linebreak
        customised database \linebreak
        order system, ticket system, online form \linebreak
        wiki \linebreak
        prefer not to answer \linebreak
        other & Multiple choice & answered “yes, a systematic overview is achieved” or “yes, but a systematic overview is not yet achieved” in Question 3 \\ \hline

        20 & How reliable do you consider the collection of publication cost data for the following sources of funding at your institution? \linebreak
        20.1: funds managed by a central service facility (e.g. via a publication fund) \linebreak
        20.2: third-party funding acquired by research and teaching departments (e. g. faculty, institute, department, section) \linebreak
        20.3: budgetary resources allocated to research and teaching departments (e. g. faculty, institute, department, section) & 
        very reliable \linebreak
        reliable \linebreak
        fair \linebreak
        unreliable \linebreak
        very unreliable \linebreak
        prefer not to answer & matrix & answered “yes, a systematic overview is achieved” or “yes, but a systematic overview is not yet achieved” in Question 3 \\ \hline

        21 & Does your institution pursue the continuous collection of publication costs during the reporting year? & 
        yes, this is already achieved \linebreak
        yes, this is not yet achieved \linebreak
        no, but this is planned \linebreak
        no, this is not planned \linebreak
        prefer not to answer \linebreak
        other & single choice & answered “yes, a systematic overview is achieved” or “yes, but a systematic overview is not yet achieved” in Question 3 \\ \hline

        22 & Are the collected publication costs used for strategic decisions at your institution (e.g. as part of resource planning)? & 
        yes \linebreak
        no \linebreak
        prefer not to answer \linebreak
        other & single choice & answered “yes, a systematic overview is achieved” or “yes, but a systematic overview is not yet achieved” in Question 3 \\ \hline

        23 & How do you evaluate the contribution of collecting publication costs to shaping the open access transformation overall? & 
        very important \linebreak
        important \linebreak
        moderately important \linebreak
        slightly important \linebreak
        unimportant \linebreak
        prefer not to answer & single choice & none \\ \hline

        24 & How realistic do you think it is to implement an information budget at your institution by 2025? & 
        this is already implemented \linebreak
        very realistic \linebreak
        rather realistic \linebreak
        maybe \linebreak
        rather unrealistic \linebreak
        very unrealistic \linebreak
        prefer not to answer & single choice & none \\ \hline

        25 & Do you have any additional remarks on collecting publication costs at your institution that you would like to share with us? & 
        NA & free text & none \\ \hline

        26 & Would you like to be informed when results of this survey are published? & 
        yes (please enter your email address in the comment) \linebreak
        no & single choice & none \\ \hline
\end{longtable}

\newpage

\section*{References}
\printbibliography[heading=none]

\end{document}